\documentclass[11pt]{article}
\usepackage{geometry}                % See geometry.pdf to learn the layout options. There are lots.
\usepackage{graphicx}
\usepackage{amssymb}
\usepackage{epstopdf}
\usepackage{amsmath}
\begin{document}

\title{Comments on: A Universe from Nothing}
\author{Ikjyot Singh Kohli \\isk@mathstat.yorku.ca \\York University - Department of Mathematics and Statistics \\ Toronto, Ontario}
%\address{Department of Mathematics and Statistics, York University}
%\email{isk@mathstat.yorku.ca}
\date{May 30, 2015}                                           % Activate to display a given date or no date

\maketitle

\begin{abstract}
We study some claims in Krauss' recent book, \emph{A Universe from Nothing: Why there is something rather than nothing}, that are employed to show that a universe can come from ``nothing''. In this brief paper, we show that many of the claims are not supported in full by modern general relativity theory or quantum field theory in curved spacetime.
\end{abstract}

The purpose of this paper is to look at in detail some of the arguments put forward in Lawrence Krauss' book, \emph{A Universe from Nothing: Why There is Something Rather Than Nothing}. Krauss puts forward arguments that are claimed to be based on general relativity and quantum field theory to show to a universe can spontaneously appear from ``nothing''. Indeed, Krauss' main claim is that ``in quantum gravity, universes can, and indeed always will, spontaneously appear from nothing''\cite{krauss},  which he bases on the Wheeler-DeWitt (WDW) version of quantum gravity. In fact, Krauss explicitly cites two formalisms built upon the WDW approach to quantum gravity, the Hartle-Hawking no boundary proposal \cite{harlehawking}, and Vilenkin's  tunnelling idea \cite{vilenkin}, both of which, as we describe in this paper have serious issues, and certainly cannot be considered as leading to universes \emph{from nothing}.

It should be noted that a universe being formed from nothing is not a new concept. Tryon \cite{tryon} was the first to suggest that the universe may have arisen from nothing, Vilenkin \cite{vilenkin} \cite{vilenk3} \cite{vilenk4} \cite{vilenk5} has done considerable work in trying to demonstrate that the universes can come from ``nothing''. Zeldovich and Starobinskii \cite{zeld1} described a quantum creation scenario of a universe with non-trivial topology. Most recently, He, Gao and Cai \cite{hegaocai} suggested yet another proposal for the spontaneous creation of a universe from nothing based on the WDW formalism. Lapiedra and Morales-Lladosa \cite{lapiedra} suggested a newer proposal in which a closed de Sitter universe could be created from nothing.

The issue with all of these works is that on one hand, the authors claim that these proposals are universes \emph{from nothing}, while on the other hand assume at a minimum all of the complex machinery of variational principles, differential and pseudo-Riemannian geometry, topology, quantum field theory, and general relativity, while \emph{never} addressing the deeper issue of where the latter come from. The other issue is that if all of this machinery is supposed to create universes from ``nothing'', which is clearly not nothing in any sense of the word, then why do the authors go to such lengths to describe this machinery as nothing? We believe that this is at the core of these issues. Is it due to some type of philosophical bias? A much deeper discussion on the philosophy of cosmology is needed to seriously address these issues. Ellis \cite{ellisphil} has done some considerable work in addressing some of these issues.

In this paper, we are particularly concerned with the claims made in Krauss' book, and will attempt to address them and show that the claims made in fact are widely not supported by general relativity or quantum field theory in curved spacetime. 

As mentioned in the introductory paragraph, Krauss' main claim is that ``in quantum gravity, universes can, and indeed always will, spontaneously appear from nothing'', which he bases on the Wheeler-DeWitt (WDW) version of quantum gravity. In fact, Krauss explicitly cites two formalisms built upon the WDW approach to quantum gravity, the Hartle-Hawking no boundary proposal \cite{harlehawking}, and Vilenkin's  tunnelling idea \cite{vilenkin}. 

Following \cite{elliscosmo} (and references therein), we note that in the WDW formalism, one represents the quantum state of the universe as $\Psi(h_{ij})$ on superspace, where $h_{ij}$ is a spatial metric, which itself is subject to the Hamiltonian constraint.  In the literature, the Wheeler-DeWitt approach has only been applied to minisuperspaces. Superspace is the space of all spatial metrics, and each point in superspace corresponds to a spatial metric $h_{ab}$. DeWitt's supermetric is defined as \cite{hervik}
\begin{equation}
G^{abij} = \frac{1}{4} \sqrt{h} \left(h^{ai}h^{bj} + h^{aj}h^{bi} - 2h^{ab} h^{ij}\right).
\end{equation}
One obtains a minisuperspace model by working with universe models that have finite degrees of freedom such as the FLRW or Bianchi models. With the aforementioned wave functional, the Wheeler-DeWitt equation takes the form (in the case of minisuperspace models)
\begin{equation}
\mathcal{H} \Psi[h_{ij}] = 0,
\end{equation}
and to solve such an equation, one needs conditions on the spatial geometry usually given in terms of boundary conditions on $\Psi$. One typically solves the Wheeler-DeWitt equation using a path integral formulation. In the Hartle-Hawking no-boundary proposal \cite{harlehawking} which is also mentioned in Krauss' book, we have
\begin{equation}
\Psi[h_{ij}] = \int \mathcal{D}g_{uv} \exp(-I[g_{uv}]),
\end{equation}
where $\mathcal{D}g_{uv}$ is a measure on the space of 3-geometries, and $I[g_{uv}]$ is the Euclidean action, which has a $S^{3}$ geometry as its boundary.  This methodology leads to a ``beginning of time'', where a classical description then becomes valid.

On the other hand, Vilenkin's  method \cite{vilenkin} produced a variant of the Wheeler-DeWitt equation for FLRW universes as
\begin{equation}
\Psi''(a) - \left[a^2 - a_{1}^{-2} a^{4}\right]\Psi(a) = 0.
\end{equation}
This equation gives tunnelling probabilities for the wave function ``from nothing'' to a closed universe of radius $a_{1}$. 

These methods are essentially the quantum gravity approaches Krauss refers to to show that it is plausible that a universe can come from nothing, but the nothing that Krauss refers to is in fact no space and no time. We do not agree that the Wheeler-DeWitt approach entails a universe coming from no space for the reason that the entire DeWitt formalism relies on an underlying superspace, which as we mentioned above, is the space of all spatial metrics, $h_{ij}$. Since both approaches above are Wheeler-DeWitt equations, they also exist on some space, namely, this superspace/minisuperspace. In particular, for such a proposal to be considered as a valid physics-based proposal, it has to be at least in principle, testable. Namely, one would have to show that preceding the big bang, or the creation of our universe, that there really was such a superspace in existence. It is not clear how at the present time that we could even begin to consider how this could be accomplished.

Notwithstanding the previous point, there are significant problems with both approaches, many types of divergences occur, namely that the path integral itself is ultraviolet divergent, and in fact, cannot be renormalized. In the Hartle-Hawking no-boundary proposal in particular, conformal modes lead to the Einstein-Hilbert action not being bounded from below, which in turn implies that the sum over all 4-geometries leads to a sum over topologies that cannot be computed \cite{donpage1}. 

In Vilenkin's approach, there is also a problem of time. There are some approaches that try to treat $a$ in the Vilenkin equation as an effective time variable, but, as Barbour \cite{Barbour} has pointed out, it is very difficult to make this work from a practical sense. One essentially has from these approaches that $\dot{|\Psi(t)\rangle} = 0$, which implies a static solution, and the concept of a time-evolving universe is thus difficult to see. Another important point is how exactly one interprets the concept of a ``wavefunction'' and probabilities when there is only \emph{one} object. Can one even give any meaningful definition of the wavefunction of a universe in these contexts? 

In Krauss' book, the concept of superspace is not mentioned a single time, even though this is the entire geometric structure for which the proposal he is putting forward of a universe coming from nothing is based upon. There is also a deep philosophical issue that \emph{cannot} be ignored. Superspace is the space of all possible 3-geometries, and the question is, what types of universes should be considered as part of a particular superspace. For example, one can consider the Bianchi cosmologies which are spatially homogeneous and anisotropic cosmological models, which have three degrees of freedom in the mini-superspace sense, of which the FLRW cosmologies are special cases. The existence of this structure is not predicted by the WDW formalism, it is \emph{assumed} to exist, which itself goes back to the theme of this paper. How can one claim that something arises from nothing, when this nothing is at minimum, minisuperspace? 

Much of the motivation of universes spontaneously being emitted from nothing seems to be the phenomenon that Krauss describes as particles being spontaneously emitted from a vacuum state. This in itself is quite problematic, as it is based on the naive particle interpretation of quantum field theory. It is well understood that in the context of cosmological models and more general spacetimes, one simply does not have a time-translation symmetry because of a lack of timelike Killing vector field. Therefore, the very definition of particles is undefined for general curved spacetimes, and only defined in the context of general relativity for asymptotically flat spacetimes \cite{waldqft}. The reason is that for the particle interpretation to work, one needs to be able to decompose the quantum field into positive and negative frequency parts, which in itself depends heavily on the presence of a such a time translation symmetry in either an asymptotically flat spacetime or a Minkowski spacetime.  The problem of course is that, our universe, or any spatially homogeneous and non-static universe, that is, one that does not contain a global timelike Killing vector field is necessarily \emph{not} asymptotically flat. This can be seen from the arguments given in \cite{stephani}. Namely, consider a spacetime $(\hat{M}, \hat{g}_{ab})$. Let this spacetime have the following three properties:
\begin{itemize}
\item There exists a function $\omega \geq 0 \in C^3$, such that $g_{ab} = \omega^2 \hat{g}_{ab}$,
\item on the boundary $\omega = 0$, and $\omega_{,a}  \neq 0$,
\item Every null geodesic intersects the boundary in two points.
\end{itemize}
These spacetimes are called asymptotically simple. However, if we now associate the metric tensor, $g_{ab}$  with the Einstein field equations, the existence of these three conditions implies that the spacetime is asymptotically flat. It is only under these three conditions, for which one can in a meaningful way talk about ``particles''. The idea of these particles being spontaneously emitted from a vacuum is also not correct for the following reason. Following \cite{waldqft}, we will consider a two-level quantum mechanical system which is coupled to a Klein-Gordon field, $\phi$ in a Minkowski spacetime, for simplicity. The combined system will have a total Hamiltonian of the form
\begin{equation}
\label{ham1}
\mathcal{H} = \mathcal{H}_{\phi} + \mathcal{H}_{q} + \mathcal{H}_{int},
\end{equation}
where $\mathcal{H}_{\phi}$ is the Hamiltonian of the free Klein-Gordon field. We will consider the quantum mechanical system to be an unperturbed two-level system with energy eigenstates $| x_{o} \rangle$ and $|x_{1} \rangle$, with energies $0$ and $\epsilon$ respectively, so we can define
\begin{equation}
\label{ham2}
\mathcal{H}_{q} = \epsilon \hat{A}^{\dagger} \hat{A},
\end{equation}
where we define
\begin{equation}
\hat{A} |x_{0} \rangle = 0, \quad \hat{A} |x_{1} \rangle = |x_{0} \rangle.
\end{equation}
The interaction Hamiltonian is defined as
\begin{equation}
\mathcal{H}_{int} = e(t) \int \hat{\psi}(\mathbf{x}) \left(F(\mathbf{x}) \hat{A} + o\right) d^{3}x,
\end{equation}
where $F(\mathbf{x})$ is a spatial function that is continuously differentiable on $\mathbb{R}^{3}$ and $o$ denotes the Hermitian conjugate. One then calculates to lowest order in $e$, the transitions of a two-level system. In the interaction picture, denoting $\hat{A}_{s}$ as the Schrodinger picture operator, one obtains
\begin{equation}
\hat{A}_{I}(t) = \exp(-i \epsilon t) \hat{A}_{s}.
\end{equation}
Therefore, we have that
\begin{equation}
(\mathcal{H}_{int})_{I} = \int \left(e(t) \exp(-i \epsilon t) F(\mathbf{x}) \psi_{I}(t,\mathbf{x}) \hat{A}_{s} + o\right) d^{3}x. 
\end{equation}
Using Fock space index notion, we can then consider for some $\Psi \in \mathbb{H}$, where $\mathbb{H}$ is the associated Hilbert space, and note that the field is in the state
\begin{equation}
|n_{\Psi} \rangle = \left(0, \ldots, 0, \Psi^{a_{1}} \ldots \Psi^{a_{n}}, 0, \ldots \right).
\end{equation}
The initial state of the full system is then given by
\begin{equation}
|\Psi_{i} \rangle = | x \rangle |n_{\Psi} \rangle.
\end{equation}
One then obtains the final state of the system as being
\begin{equation}
\label{3317}
|\Psi_{f} \rangle = |n _{\Psi} \rangle |x \rangle + \sqrt{n+1} \| \lambda \| (\hat{A} |x \rangle) |(n+1)^{'}\rangle - \sqrt{n} (\lambda, \Psi) (\hat{A}^{\dagger} |x\rangle) |(n-1)_{\Psi}\rangle,
\end{equation}
where
$| (n+1)^{'} \rangle$ is defined as in Eq. (3.3.18) in \cite{waldqft}, 
and $\lambda$ is defined as in Eq. (3.3.15) in \cite{waldqft}.

The key point is that if $|x \rangle = |x_{0} \rangle$, that is, the system is in its ground state, Eq. \eqref{3317} shows explicitly that this two-level system can make a transition to an excited state, and vice-versa. Note that the probability of making a downward transition is proportional to $(n+1)$, and even when $n = 0$, this probability is non-zero. This in the \emph{particle interpretation} is interpreted as saying that the quantum mechanical system can spontaneously emit a particle. However, the above calculation in deriving Eq. \eqref{3317} explicitly shows that it is the interaction of the quantum mechanical system with the quantum field that is responsible for the so-called spontaneous particle emission. This misleading picture of the vacuum state is precisely promoted by the particle interpretation of quantum field theory. As the work above shows, this is not spontaneous particle emission from ``nothing'' in any sense of the word. One must have both a well-defined quantum mechanical system interacting with a well-defined vacuum state for such spontaneous emission to occur, we emphasize that these are not nothing! 

Finally, Krauss makes the very problematic claim that ``the structures we can see, like stars and galaxies, were all created by quantum fluctuations from nothing'' \cite{krauss}. The issue that is not being addressed here is the \emph{fundamental} transition problem of going from a pure quantum state to a classical state. The reason for this is first and foremost due to the problem of quantization, which in its general form, is seldom addressed in even advanced physics textbooks on quantum mechanics. Following \cite{abrahammarsden}, let us denote the classical phase space as $Q = \mathbb{R}^{n}$. Then a full quantization of $Q$ is a map taking classical observables $f:(q,p) \in T^{*}\mathbb{R}^{n} \to \hat{f} \in \mathbb{H}$, where $\hat{f}$ denotes self-adjoint operators on the Hilbert space $\mathbb{H}$, such that:
\begin{enumerate}
\item $\hat{(f+g)} = \hat{f} + \hat{g}$,
\item $\hat{(\lambda f)} = \lambda \hat{f}, \quad \lambda \in \mathbb{R}$,
\item $  \hat{\{f,g\}} = (1/i) [\hat{f},\hat{g}]$,
\item $ \hat{\mathbf{1}} = \hat{I}$, ($\mathbf{1}$ is the constant function, while $\hat{I}$ denotes the identity),
\item $\hat{q}^{i}, \hat{p}_{j}$ act irreducibly on $\mathbb{H}$.
\end{enumerate}
Note that the point of the last requirement is to really allow one to take $\mathbb{H} = L^2(\mathbb{R}^{n})$, so that $\hat{p}_{j} = (1/i) \partial / \partial q_{j}$, which defines the Schrodinger representation. One also must include the following modified condition to allow for spin in which the position and momentum operators are represented by a direct sum of finitely many copies of the Schrodinger representation, such that
\begin{equation}
\hat{q}_{i} \phi(x) = q_{i} \phi(x), \quad \hat{p}_{j}\phi(x) = -i \frac{\partial \phi(x)}{\partial x_{j}}.
\end{equation}
Groenwold \cite{groenwold} and van Hove \cite{vanhove} explicitly showed that no quantization simultaneously satisfying all of these properties is possible. In fact, as described in detail one can only simultaneously satisfy (1)-(4) above, but not (5) or its modification, which in itself only works in pre-quantization. This goes to the broader point made by Groenwold \cite{groenwold}, that the quantization rule due to Dirac
\begin{equation}
\left\{A,B\right\} \to \frac{1}{i\hbar} \left[\hat{A},\hat{B}\right],
\end{equation}
is not true in general. In fact, let $[\hat{q},\hat{p}] = i \hbar$, then from
\begin{equation}
\left\{q^3, p^3\right\} + \frac{1}{12} \left\{ \left\{p^2, q^3\right\}, \left\{q^2, p^3\right\}    \right\} = 0,
\end{equation}
we should obtain that
\begin{equation}
\frac{1}{ih} \left[\hat{q}^{3}, \hat{p}^3\right] + \frac{1}{12 i \hbar} \left[\frac{1}{i\hbar} \left[\hat{p}^{2}, \hat{q}^3\right], \frac{1}{i\hbar} \left[\hat{q}^{2}, \hat{p}^3\right]  \right] = 0,
\end{equation}
but in fact, as Groenwold showed the latter equation evaluates to $-3\hbar^2$! 

Therefore, the point that Krauss makes that all classical structures arise from quantum fluctuations is not true in general, as it is not even possible to fully quantize all possible classical systems. Further, with respect to the transition problem, one needs much more additional structure than the Hilbert space of quantum mechanics to get back to classical mechanics. This is easily seen from the fact that the spaces $L^2(\mathbb{R}^{n})$ for different values of $n$ are isomorphic. One would need to introduce a distinguished class of operators, or a well-defined set of commuting operators to obtain the classical system. This goes back to the philosophical problem of what one means by nothing. Indeed, for Krauss' claim to hold true, this entire quantization structure including these sets of operators must also be somehow defined, which, as we point out again, Krauss has no plausible explanation of the origins of these necessary structures.

Further, following \cite{ellisctns}, it is important to note that in the \emph{real} universe, the future evolution is not uniquely predicted by the past, precisely because of inflationary perturbations! The inhomogeneities that occurred on the last scattering surface were the result of quantum fluctuations during inflation. They were not determined uniquely by the state of the universe at the start of inflation because of the inherent quantum uncertainties. As is also pointed out in \cite{ellisctns}, suppose we knew every detail of the state of the Earth and the life on it two billion years ago, this would not uniquely predict that humans would exist today, because the random quantum events leading to cosmic ray emission can change the genetic traits of animals, thus influencing biological evolution. 

Quantum fluctuations are also not fully responsible for allowing emergence to occur, and there must be genuine emergence, namely, the emergence of complexity that that occurs in order for the macroscopic levels or order that we see in the universe. For this to occur, there must be a precise interplay between not only bottom-up causation, but top-down causation as well. The interested reader is encouraged to see \cite{topdown1} for further details on this very important point.

Although what is presented above is our main rebuttal to the arguments that Krauss presents, in what follows we briefly touch on some other points in Krauss' book that are also not correct. 

Krauss claims that ``Considering the geometry of the universe is like imagining a pencil balancing vertically on its point on a table. The slightest imbalance one way or the other and it will quickly topple. So it is for a flat universe. The slightest departure from flatness quickly grows. Thus, how could the universe be so close to being flat today if it were not exactly flat.'' 
The entire basis of this claim is that the universe was always isotropic and therefore, spatially homogeneous. Therefore, the implication is that even in the asymptotic past, the universe was of FLRW-type. In this case, we can see the validity of Krauss' claim by considering the following. Let the metric tensor of the FLRW be defined as in \cite{ellis}:
\begin{equation}
ds^2 = -dt^2 + l^2(t)\left(dr^2 + f^2(r)(d\theta^2 + \sin^2 \theta d \phi^2)\right),
\end{equation}
where $l$ is the typical length-scale function, and $f(r) = \sin r, r, \sinh r$ depending on whether the FLRW model being considered is closed, flat, or open, i.e,. $k = +1, 0,$ or $-1$. Let us define 
\begin{equation}
l = l_{0} e^{\tau},
\end{equation}
where $\tau$ is a dimensionless time variable, and further define the quantity
\begin{equation}
\frac{dt}{d\tau} = \frac{1}{H},
\end{equation}
where $H$ is the Hubble parameter, which leads to 
\begin{equation}
\frac{dH}{d\tau} = -(1+q)H,
\end{equation}
where 
\begin{equation}
q =  \frac{1}{2}\left(3\gamma - 2\right) \Omega.
\end{equation}
The standard Bianchi identities then imply that
\begin{equation}
\frac{d \Omega}{d \tau} = -(3\gamma -2)(1-\Omega)\Omega.
\end{equation}
This equation describes the evolution of \emph{all} single-fluid FLRW models with linear equation of state. The fixed points of this equation are given by $\Omega = 0$ and $\Omega = 1$, with the latter corresponding to the flat FLRW universe. As shown in \cite{ellis}, upon analyzing the phase space defined by $\Omega \geq 0$, the point $\Omega = 1$ is source of the system if $\gamma > \frac{2}{3}$ and a sink of the system if $\gamma < \frac{2}{3}$. Indeed, $\Omega = 1$ is a saddle if and only if $\gamma = 2/3$.  Therefore, as one can see, Krauss' claim only holds true for when the equation of state parameter $\gamma$ is such that $\gamma \geq \frac{2}{3}$. 

However, it has been suggested that the early universe being hot, and dense, may have contained anisotropic matter usually modelled by shear viscosity in the energy-momentum tensor \cite{hervik}. In addition, primordial magnetic fields have also been suggested to exist in the early universe \cite{andokusenko} \cite{grassorub}, \cite{gregorietal}. In this case, the presence of magnetic fields introduces a shear component in the energy-momentum tensor as well, so the FLRW models would not be relevant at this time in the universe's evolution. One must therefore, at a minimum, consider anisotropic cosmologies. One can in principle, still maintain the assumption of spatial homogeneity, thus working with the Bianchi cosmological models. Indeed, much work has been done in this area \cite{hughstonjacobs}, \cite{leblanc1}, \cite{leblanc2}, \cite{collins}, \cite{leblanc3}, \cite{isk2}. In the latter papers, it is shown that a flat FLRW universe is a local sink of the Einstein field equations, and further analysis of the global behaviour of the orbits show that under certain reasonable conditions, a flat FLRW universe is a global stable asymptotic state of the system. In this case, deviations from the flat FLRW point will not change the trajectory of the system. This can only occur if there are bifurcations in the parameter space, which is an entirely different matter than Krauss is claiming. Indeed, the dynamics of the Bianchi models can be obtained in an intuitive way by considering the orthonormal frame formalism \cite{ellismac}, and then employing expansion-normalized variables approach \cite{hewittbridsonwainwright} to obtain the following form of the Einstein field equations:
\begin{eqnarray}
\label{eq:evolutionsys1}
\Sigma_{ij}' &=& -(2-q)\Sigma_{ij} + 2\epsilon^{km}_{(i}\Sigma_{j)k}R_{m} - \mathcal{S}_{ij} + \Pi_{ij} \nonumber \\
N_{ij}' &=& qN_{ij} + 2\Sigma_{(i}^{k}N_{j)k} + 2 \epsilon^{km}_{(i}N_{j)k}R_{m} \nonumber \\
A_{i}' &=& qA_{i} - \Sigma^{j}_{i}A_{j} + \epsilon_{i}^{km}A_{k} R_{m}\nonumber \\
\Omega' &=& (2q - 1)\Omega - 3P - \frac{1}{3}\Sigma^{j}_{i}\Pi^{i}_{j} + \frac{2}{3}A_{i}Q^{i} \nonumber \\
Q_{i}' &=& 2(q-1)Q_{i} - \Sigma_{i}^{j}Q_{j} - \epsilon_{i}^{km}R_{k}Q_{m} + 3A^{j}\Pi_{ij} + \epsilon_{i}^{km}N_{k}^{j}\Pi_{jm}.
\end{eqnarray}
These equations are subject to the constraints
\begin{eqnarray}
\label{eq:constraints1}
N_{i}^{j}A_{j} &=& 0 \nonumber \\
\Omega &=& 1 - \Sigma^2 - K \nonumber \\
Q_{i} &=& 3\Sigma_{i}^{k} A_{k} - \epsilon_{i}^{km}\Sigma^{j}_{k}N_{jm}.
\end{eqnarray}
In Eqs. (\ref{eq:evolutionsys1}) and (\ref{eq:constraints1}) we have made use of the following notation:
\begin{equation}
\label{eq:notation1}
\left(\Sigma_{ij}, R^{i}, N^{ij}, A_{i}\right) = \frac{1}{H}\left(\sigma_{ij}, \Omega^{i}, n^{ij}, a_{i}\right) , \quad \left(\Omega, P, Q_{i}, \Pi_{ij}\right) = \frac{1}{3H^2}\left(\mu, p, q_{i}, \pi_{ij}\right).
\end{equation}
In the expansion-normalized approach, $\Sigma_{ab}$ denotes the kinematic shear tensor, and describes the anisotropy in the Hubble flow, $A_{i}$ and $N^{ij}$ describe the spatial curvature, while $\Omega^{i}$ describes the relative orientation of the shear and spatial curvature eigenframes.
In addition,  $\mu$ and $p$ denote the \emph{total} energy density and total effective pressure, and are found by evaluating
\begin{equation}
\label{eq:totaldensitypressuredef}
\mu = u^{a} u^{b} T_{ab}, \quad p = \frac{1}{3} h^{ab} T_{ab},
\end{equation}
where, $h_{ab} = u_{a}u_{b} + g_{ab}$ denotes the projection tensor, and $u^{a}$, the fluid four-velocity \cite{herviklim}.
It can then be shown by generalizing the equations Eqs. \eqref{eq:evolutionsys1} to different Bianchi types that a flat FLRW universe given by
\begin{equation}
\left(\Sigma_{+}, \Sigma_{-}, N_{i}, A_{i}\right) = \left(0,0,0,0\right)
\end{equation}
is a local sink of this dynamical system under certain conditions of the equation of state parameter (and possibly other phenomenological parameters). In addition, for certain Bianchi types, one can prove that the FLRW flat universe point is globally stable. For further details, the interested reader should consult Chapters 6 and 7 of \cite{ellis}.

Further, Krauss also claims on several occasions that the total energy of a closed universe is zero. The issues with this claim are as follows. 

As stated on page 457 in \cite{mtw}, ``for a closed universe the total mass-energy and angular momentum are undefined and undefinable''.  In addition, following \cite{baumgarte}, we note that the total mass-energy of a system in general relativity cannot be generally defined. There are however, a few tools one can employ to measure the total mass-energy of a system in the case of asymptotically flat spacetimes. The first is the ADM mass, defined by:
\begin{equation}
M_{ADM} = \frac{1}{16 \pi} \int_{\partial \Sigma_{\infty}} \sqrt{\gamma} \gamma^{jn} \gamma^{im} \left(\gamma_{mn,j} - \gamma_{jn,m}\right) dS_{i},
\end{equation}
which requires the space-time to be asymptotically flat.  Another tool is the Komar mass, 
\begin{equation}
M_{K} = \frac{1}{4 \pi} \int_{\Sigma} d^3x \sqrt{\gamma} n_{a} J^{a}_{(t)},
\end{equation}
which also requires the spacetime to have an asymptotically flat region. The problem of of course is that, our universe, or any spatially homogeneous and non-static, that is, one that does not contain a global timelike Killing vector is necessarily \emph{not} asymptotically flat.

In conclusion, we point out that based on the arguments presented in this paper, the main claims in Krauss' book are not supported by General Relativity, Relativistic Cosmology, or Quantum Field Theory in curved spacetime. Namely, Krauss' ``nothing'' on some occasions is the quantum vacuum, and on other occasions, it is the minisuperspace of the WDW formalism. In either case, Krauss does not bother to address the underlying laws of physics/mathematics that govern these principles, nor does he bother to point out that none of his claims are testable, which is absolutely key for a theory to be considered scientific. For these reasons, we strongly believe that Krauss' book is largely motivated by a particular set of philosophical ideologies of which principles from science are being extrapolated at best to support these arguments, which by the end of the day, the reader hopefully will realize are filled with all sorts of gaping holes and incorrect notions.

\newpage
\bibliographystyle{ieeetr} 
\bibliography{sources}

\begin{thebibliography}{10}

\bibitem{krauss}
L.~M. Krauss, {\em A Universe from Nothing: Why There is Something Rather than
  Nothing}.
\newblock Atria Books, kindle~ed., 2012.

\bibitem{harlehawking}
J.~Hartle and S.~Hawking, ``Wave function of the universe,'' {\em Phys. Rev.
  D}, vol.~28, p.~2960, 1983.

\bibitem{vilenkin}
A.~Vilenkin, ``Creation of universes from nothing,'' {\em Physics Letters B},
  vol.~117, pp.~25--28, 1982.

\bibitem{tryon}
E.~P. Tryon, ``Is the universe a vacuum fluctuation?,'' {\em Nature}, vol.~246,
  pp.~396--397, 1973.

\bibitem{vilenk3}
A.~Vilenkin, ``Quantum creation of universes,'' {\em Physical Review D},
  vol.~30, pp.~509--511, 1984.

\bibitem{vilenk4}
A.~Vilenkin, ``Quantum origin of the universe,'' {\em Nuclear Physics B},
  vol.~252, pp.~141--152, 1985.

\bibitem{vilenk5}
A.~Vilenkin, ``Quantum cosmology and the initial state of the universe,'' {\em
  Physical Review D}, vol.~37, pp.~888--897, 1988.

\bibitem{zeld1}
Y.~Zeldovich and A.~Starobinskii, ``Quantum creation of a universe with
  nontrivial topology,'' {\em Soviet Astronomy Letters}, vol.~10, pp.~135--137,
  1984.

\bibitem{hegaocai}
D.~He, D.~Gao, and Q.-y. Cai, ``Spontaneous creation of the universe from
  nothing,'' {\em Physical Review D}, vol.~89, p.~083510, 2014.

\bibitem{lapiedra}
R.~Lapiedra and J.~A. Morales-Lladosa, ``On cosmic quantum tunneling from
  ``nothing'','' {\em Journal of Physics: Conference Series}, vol.~600,
  p.~012020, 2015.

\bibitem{ellisphil}
G.~F. Ellis, ``Issues in the philosophy of cosmology,'' {\em
  arXiv:astro-ph/0602280}, 2006.

\bibitem{elliscosmo}
G.~F. Ellis, R.~Maartens, and M.~A. MacCallum, {\em Relativistic Cosmology}.
\newblock Cambridge University Press, first~ed., 2012.

\bibitem{hervik}
{\O}.~Gr{\o}n and S.~Hervik, {\em Einstein's General Theory of Relativity: With
  Modern Applications in Cosmology}.
\newblock Springer, first~ed., 2007.

\bibitem{donpage1}
D.~N. Page, ``Boundary conditions and predictions of quantum cosmology,'' {\em
  Invited lecture for Session COT5 of the 11th Marcel Grossmann Meeting on
  General Relativity}, 2006.

\bibitem{Barbour}
J.~Barbour, {\em The End of Time: The Next Revolution in Physics}.
\newblock Oxford University Press, first~ed., 2001.

\bibitem{waldqft}
R.~M. Wald, {\em Quantum Field Theory in Curved Spacetime and Black Hole
  Thermodynamics}.
\newblock University of Chicago Press, first~ed., 1994.

\bibitem{stephani}
H.~Stephani, {\em Relativity: An Introduction to Special and General
  Relativity}.
\newblock Cambridge University Press, third~ed., 2004.

\bibitem{abrahammarsden}
R.~Abraham and J.~E. Marsden, {\em Foundations of Mechanics}.
\newblock AMS Chelsea Publishing, second~ed., 1978.

\bibitem{groenwold}
H.~Groenwold, ``On the principles of elementary quantum mechanics,'' {\em
  Physica}, vol.~12, pp.~405--460, 1946.

\bibitem{vanhove}
L.~Van~Hove, ``Sur certaines representations unitaires d'un groupe infini de
  transformations,'' {\em Mem. de l'Acad Roy. de Belgique}, vol.~{{XXVI}},
  pp.~61--102, 1951.

\bibitem{ellisctns}
G.~F. Ellis, ``Necessity, purpose, and chance: The role of randomness and
  indeterminism in nature from complex macrosystems to relativistic
  cosmology,''

\bibitem{topdown1}
G.~F. Ellis, ``Top-down causation and emergence: some comments on mechanisms,''
  {\em Interface Focus}, vol.~2, pp.~126--140, 2012.

\bibitem{ellis}
J.~Wainwright and G.~Ellis, {\em Dynamical Systems in Cosmology}.
\newblock Cambridge University Press, first~ed., 1997.

\bibitem{andokusenko}
S.~Ando and A.~Kusenko, ``Evidence for gamma-ray halos around active galactic
  nuclei and the first measurement of intergalactic magnetic fields,'' {\em The
  Astrophysical Journal Letters}, vol.~722, pp.~L39--L44, 2010.

\bibitem{grassorub}
D.~Grasso and H.~R. Rubinstein, ``Magnetic fields in the early universe,'' {\em
  Physics Reports}, vol.~348, pp.~163--266, 2001.

\bibitem{gregorietal}
G.~Gregori and et~al., ``Generation of scaled protogalactic seed magnetic
  fields in laser-produced shock waves,'' {\em Nature}, vol.~481, pp.~480--483,
  2012.

\bibitem{hughstonjacobs}
L.~P. Hughston and K.~C. Jacobs, ``Homogeneous electromagnetic and
  massive-vector fields in bianchi cosmologies,'' {\em Astrophysical Journal},
  vol.~160, pp.~147--152, 1970.

\bibitem{leblanc1}
V.~LeBlanc, ``Bianchi ii magnetic cosmologies,'' {\em Classical and Quantum
  Gravity}, vol.~15, pp.~1607--1626, 1998.

\bibitem{leblanc2}
V.~LeBlanc, ``Asymptotic states of magnetic bianchi i cosmologies,'' {\em
  Classical and Quantum Gravity}, vol.~14, pp.~2281--2301, 1997.

\bibitem{collins}
C.~Collins, ``Qualitative magnetic cosmology,'' {\em Communications in
  Mathematical Physics}, vol.~27, pp.~37--43, 1972.

\bibitem{leblanc3}
V.~LeBlanc, D.~Kerr, and J.~Wainwright, ``Asymptotic states of magnetic bianchi
  $vi_{0}$ cosmologies,'' {\em Classical and Quantum Gravity}, vol.~12,
  pp.~513--541, 1995.

\bibitem{isk2}
I.~S. Kohli and M.~C. Haslam, ``Dynamical systems approach to a bianchi type i
  viscous magnetohydrodynamic model,'' {\em Phys. Rev. D}, vol.~88, p.~063518,
  Sep 2013.

\bibitem{ellismac}
G.~Ellis and M.~MacCallum, ``A class of homogeneous cosmological models,'' {\em
  Comm. Math. Phys}, vol.~12, pp.~108--141, 1969.

\bibitem{hewittbridsonwainwright}
C.~Hewitt, R.~Bridson, and J.~Wainwright, ``The asymptotic regimes of tilted
  bianchi ii cosmologies,'' {\em General Relativity and Gravitation}, vol.~33,
  pp.~65--94, 2001.

\bibitem{herviklim}
S.~Hervik, W.~C. Lim, P.~Sandin, and C.~Uggla, ``{Future asymptotics of tilted
  Bianchi type II cosmologies},'' {\em Classical and Quantum Gravity}, vol.~27,
  2010.

\bibitem{mtw}
Misner, Thorne, and Wheeler, {\em Gravitation}.
\newblock W.H. Freeman, second~ed., 1973.

\bibitem{baumgarte}
T.~W. Baumgarte and S.~L. Shapiro, {\em Numerical Relativity - Solving
  Einstein's Equations on the Computer}.
\newblock Cambridge University Press, first~ed., 2010.

\end{thebibliography}

\end{document}